\documentclass{article}
\usepackage{ijcai22}

\usepackage{times}
\usepackage{soul}
\usepackage{url}
\usepackage[hidelinks]{hyperref}
\usepackage[utf8]{inputenc}
\usepackage[small]{caption}
\usepackage{graphicx}
\usepackage{amsmath}
\usepackage{amsthm}
\usepackage{booktabs}
\usepackage{algorithm}
\usepackage{algorithmic}
\usepackage{balance}
\usepackage{xcolor}
\usepackage{tikz-qtree}
\usepackage[normalem]{ulem}

\theoremstyle{definition}
\newtheorem*{definition}{Definition}

\title{Towards Verifiable Federated Learning}

\author{
Yanci Zhang\And
Han Yu
\affiliations
School of Computer Science and Engineering, Nanyang Technological University, Singapore
\emails
\{yanci001, han.yu\}@ntu.edu.sg
}

\begin{document}

\maketitle

\begin{abstract}
  Federated learning (FL) is an emerging paradigm of collaborative machine learning that preserves user privacy while building powerful models. Nevertheless, due to the nature of open participation by self-interested entities, it needs to guard against potential misbehaviours by legitimate FL participants. FL verification techniques are promising solutions for this problem. They have been shown to effectively enhance the reliability of FL networks and help build trust among participants. Verifiable federated learning has become an emerging topic of research that has attracted significant interest from the academia and the industry alike. Currently, there is no comprehensive survey on the field of verifiable federated learning, which is interdisciplinary in nature and can be challenging for researchers to enter into. In this paper, we bridge this gap by reviewing works focusing on verifiable FL. We propose a novel taxonomy for verifiable FL covering both centralised and decentralised FL settings, summarise the commonly adopted performance evaluation approaches, and discuss promising directions towards a versatile verifiable FL framework.
\end{abstract}

\section{Introduction}
Machine learning (ML) has become a useful tool that has impacted wide-ranging applications in our daily lives. Various industries are increasingly relying on ML techniques to provide quality services based on their rich collections of data. However, traditional ML techniques require data to be located in a central facility in order to train models. In an increasingly privacy-aware world \cite{GDPR}, the privacy issues related to such a practice hinder the development of ML-empowered applications, especially in sensitive fields such as healthcare and finance. 

To enable ML techniques to continue their development trajectory under data privacy protection requirements, the federated learning paradigm \cite{konevcny2016federated} has been proposed, enabling data owners (a.k.a. FL clients or FL trainers) to train machine learning models collaboratively without compromising privacy. The classical centralised FL framework consists of a central FL server and multiple FL clients. 
The FL server first initialises the model parameters and securely passes them to the FL clients. Each client independently trains its local model using its local computing resources and local data. After this step, the client sends the model parameters back to the FL server. The server aggregates the received local models to obtain an updated global FL model, and distributes it among the clients for further training. This iterates until the model meets the convergence conditions. Apart from centralised FL, the decentralised FL architecture has also emerged. In decentralised FL, the server's role is taken up by a blockchain. Each data owner (a.k.a. trainer) performs local model training. These local model updates are shared with selected data owners (a.k.a. workers) who perform model aggregation and generate a new block containing the aggregated model on the blockchain. Information stored on the blockchain is immutable to modification and accessible by all.
 

Federated learning is a useful approach for enabling collaborative model training in a privacy-preserving manner. However, this open collaboration among potentially self-interested parties who may not know each other well has brought new challenges for FL. For example, the FL server which is responsible for selecting data owners to participate in FL, ascertaining their contributions to the model and distributing incentive payouts \cite{Zhan-et-al:2021} might deviate from the agreed protocols in pursuit of its own interest. Similarly, FL clients might also deviate from the FL protocol to free-ride on others \cite{lyu2020fppdl} or to slow down/impede the model training process \cite{Park-et-al:2021}. In addition, FL clients who forge their identities, lie about the size or quality of their datasets and local computing resources can mislead the decisions and negatively impact the FL process. Therefore, it is crucial to verify the information provided and the behaviours undertaken by the FL participants in order to build a healthy FL ecosystem. 

The research of introducing verification into FL has attracted significant attention. A verifiable FL system not only enhances the system security, but also helps build trust among legitimate participants. In centralised FL, to verify the correctness of the aggregated models produced by the FL server, the server needs to generate a proof, a signature or a ciphertext~\cite{xu2019verifynet,guo2020v,zhang2020privacy,fu2020vfl,mou2021verifiable,han2021verifiable,jiang2021pflm,madi2021secure,hahn2021versa} to prove that the aggregation result is not forged. FL clients need verification from the server in order to participate. Verification for clients' local information, including but not limited to identity and local data, is the foundation of a secure FL system~\cite{zhang2020blockchain,hahn2021versa,mou2021verifiable,lo2021blockchain,li2021efficient,zhang2021incentive}. FL client training performance is another part of client verification~\cite{kang2019incentive,zhang2020enabling,kang2020reliable,song2021reputation} as forged local model updates can lead to poor performance in the final model. Similarly, in decentralised FL, the workers need to verify the uploaded model parameters before performing model aggregation and update~\cite{kim2019blockchained,bao2019flchain,zhao2020mobile,peng2021vfchain}.


The field of verifiable federated learning is interdisciplinary as it requires expertise from machine learning, cryptography and game theory, etc. This makes it challenging for researchers new to the field to grasp the latest development. Currently, there is no survey paper on this important and rapidly developing field.
To bridge this gap, we provide a comprehensive survey of research works towards a verifiable FL framework in this paper. We propose a unique taxonomy of verifiable FL that organises existing works according to the FL system architecture, the stakeholders involved and the subjects of verification to provide readers with a multi-perspective view into this field. We analyse the limitations of current approaches, summarise the commonly adopted performance evaluation approaches, and offer promising future directions leading towards verifiable FL frameworks. 

\section{Preliminaries}

\subsection{Defining Verifiable Federated Learning}\label{st:2.1}
Before delving into our survey, we first need to define what counts as a verifiable FL approach. In~\cite{Kairouz-et-al:2021}, verifiability is defined as the ability of an FL client or the FL server to prove to others who engage in an FL protocol that it has executed the desired behaviour faithfully, without revealing the potentially private data upon which they were acting. However, this definition is strictly restricted in the centralised FL setting. Therefore, in this paper, we extend the definition of verifiability to both centralised setting and decentralised setting as follows:

\begin{definition}[Verifiable FL]
\textit{Verifiability is the ability of one party to prove to other parties in an FL protocol that it has correctly performed the intended task without deviation.}
\end{definition}


\subsection{Stakeholder Analysis}
In an FL ecosystem, there can be many model users looking for data owners to collaboratively build FL applications. To ensure trustworthy and sustainable interactions in such an open dynamic environment, there are two types of direct stakeholders for verifiable FL: 1) FL servers (only applicable in centralised FL settings), and 2) data owners. They are directly involved in the FL training processes. Indirect stakeholders include third-party auditors, researchers and developers, government agencies, policymakers, civil societies, and insurance companies, etc. Each stakeholder has unique needs for functional features from a verifiable FL system \cite{brundage2020toward}. The diverse stakeholders' needs require different verifiable FL techniques. 

FL servers need to verify that FL clients possess the resources they claim and have followed the training protocol faithfully. Data owners need to verify that the FL server is indeed protecting their data privacy. Third-party auditors need to leverage the auditable trails recorded during historical verification processes to ascertain the correctness of the verification operations. Researchers and developers need to verify whether the model performance (e.g., accuracy and running time) meets the expectation. Government agencies, policymakers and civil societies need to verify whether the FL frameworks comply with regulations. In mission-critical applications (e.g., healthcare, finance), insurance agencies need to verify if there exists any bias in the FL models involved. In order to support diverse stakeholders' needs, different verifiable FL techniques are required.

\section{The Proposed Verifiable FL Taxonomy}
\begin{figure}[ht]
\includegraphics[width=1\linewidth]{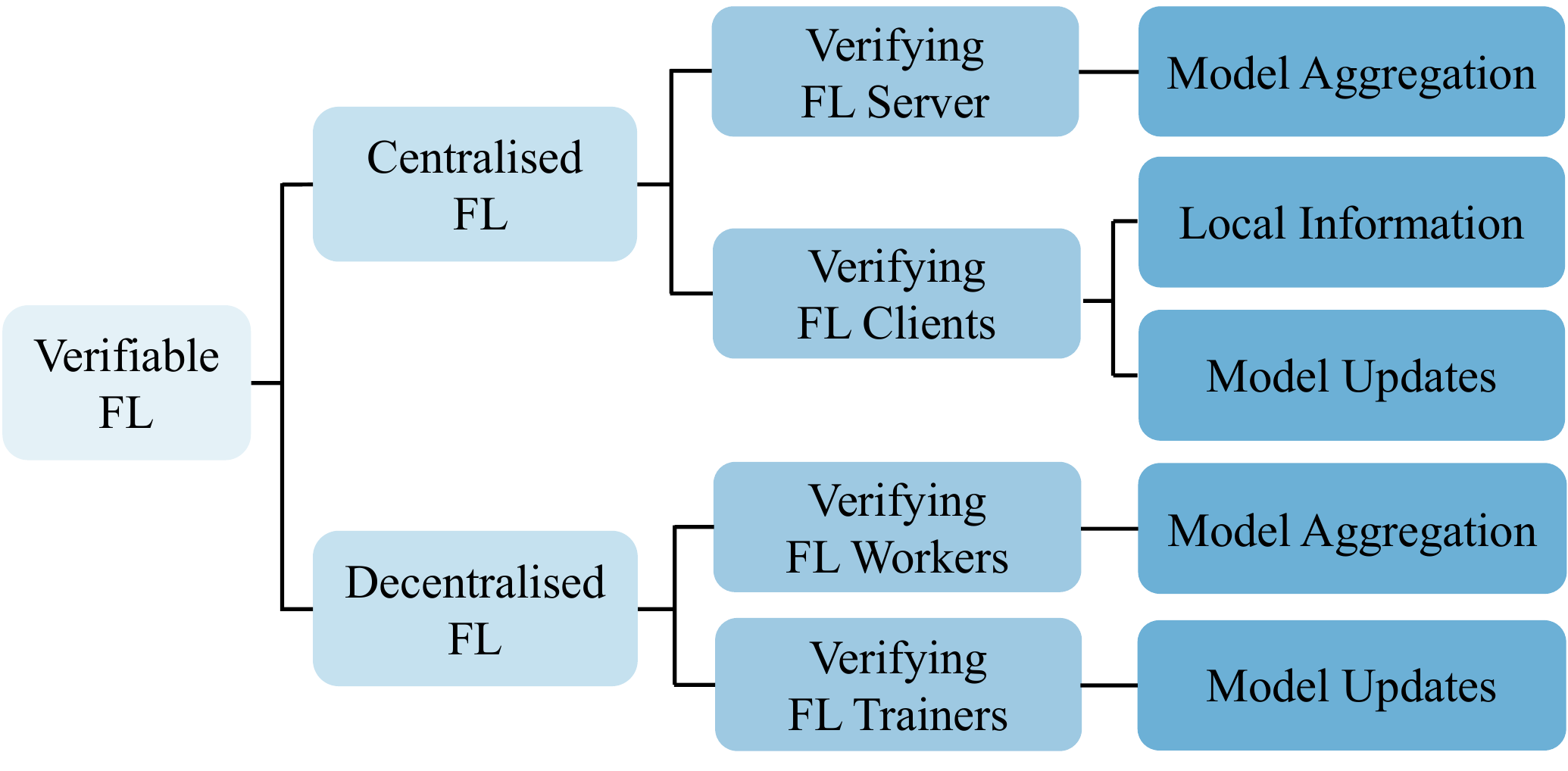}

\caption{The proposed taxonomy of verifiable FL approaches.}\label{fig:1}
\end{figure}

Based on the extended definition of FL verifiability given in Section \ref{st:2.1}, we propose a taxonomy of verifiable FL (as shown in Figure \ref{fig:1}) according to the direct stakeholders who require the given form of verification. The taxonomy first separates verifiable FL into centralised and decentralised settings. Since any legitimate FL participant can misbehave, either through not performing its tasks faithfully or maliciously exposing private information, we further classify the verifiable FL based on the target of verification. This hierarchical taxonomy provides a clear overview of the current verifiable FL research landscape.

\subsection{Verifiable Centralised FL Systems}
In centralised FL, the server and clients are the two main types of participants. Therefore, the aggregated FL model distributed by the server and the local model updates submitted by the clients are important targets of verification. 

\subsubsection{Verifying the FL Server}
The role played by the FL server during training is to aggregate local model updates submitted by FL clients and then distribute the aggregated model back to clients for further training. The reliability of the server is vital to the FL protocol. The server cannot be assumed to be always trustworthy in all scenarios \cite{Kairouz-et-al:2021}. An untruthful FL server can compromise privacy and benefits for all clients involved. Therefore, it is crucial for the clients to verify the integrity of the received aggregated FL model. 

When verifying the FL server, the server is the prover and each FL client is a verifier. In non-FL settings, SafetyNets~\cite{ghodsi2017safetynets} uses an interactive proof protocol to enable cloud service users to verify the correctness of the tasks that the cloud server performs. In each round of interaction, the user sends the cloud server a randomly selected challenge. The server proves its correctness by a short proof. Although SafetyNets is only applicable in non-FL settings and for deep neural network (DNN)-based models which can be represented as arithmetic circuits, this proof generation technique lays the foundation for FL server verification.

Existing FL server verification research commonly makes the following assumptions:
\begin{enumerate}
    \item \textit{Semi-honest FL Clients}: FL clients are assumed to truthfully upload their local model parameters, but are curious about other clients' private information. They might collude with the server to gain access to such information.
    \item \textit{Malicious FL Server}: The FL server might attempt to steal private information from clients through their local model updates. It might forge the aggregated global FL model to be distributed among the clients. 
\end{enumerate}

Under these assumptions, a number of approaches have been proposed to verify the correctness of server aggregated FL models. In general, the server is required to return the aggregated model in a certain encrypted form or with computed proof for clients to verify. The server can obtain a valid proof only by correctly executing the aggregation steps. 
Since encryption is often involved for verifying the FL server, initialisation steps (e.g., key generation) are required before model training. We divide existing works based on the entity conducting the initialisation operations.

\paragraph{Initialisation by a Trusted Third Party:}
In this setting, a trusted third party takes charge of initial key generation and assignment before training begins. This entity is assumed to be trustworthy and does not collude with any FL participant. 

VerifyNet~\cite{xu2019verifynet} is one of the first models to verify the correctness of the server returned results. Under VerifyNet, the server returns the aggregated FL model together with its proof. All clients can then verify the proof by utilising the homomorphic hash function integrated with the pseudorandom function to determine whether the aggregation is trustful or malicious. In~\cite{mou2021verifiable} and~\cite{han2021verifiable}, the authors also adopt homomorphic hash functions and pseudorandom functions to verify server aggregated results via server-generated proof. The difference between these three works mainly lies in the underlying model gradient encryption and privacy protection techniques. 

The homomorphic hash~\cite{fiore2014efficiently}, however, is computationally costly as it heavily relies on the bilinear pairing~\cite{brezing2005elliptic} which conducts the pairing operations on each entry in a vector, making the time cost linearly related to the dimension of the model gradient vectors. Therefore the communication cost is huge when the dimension of the model gradient vector is high. In view of this problem, VERIFL~\cite{guo2020v} reduces the communication overhead by requiring each client to commit only the hash of its gradient vector rather than the vector itself. In~\cite{zhang2020privacy}, the bilinear aggregate signature technique is introduced to verify the correctness of aggregated parameters. The authors leverage the Chinese Remainder Theorem to reduce communication costs. Similarly, VFL~\cite{fu2020vfl} uses a pseudorandom generator and Lagrange interpolation together with the Chinese Remainder Theorem to encrypt the client's gradient. The server performs aggregation directly using the received ciphertext. Thus, clients can verify the correctness of the aggregated model by unpacking the received aggregated ciphertext from the server. In~\cite{jiang2021pflm}, the proof returned by the server is generated by cryptographic accumulators and stored in a public blockchain for verification. Clients can apply a variant of ElGamal encryption to verify the correctness of the aggregated models. 
Initialisation by a trusted third party is a strong assumption as in real-world applications as such an entity might not always be available. 

\paragraph{Initialisation by the Server or the Client(s):} 
To suit real-world application scenarios, researchers are incorporating the initialisation step into FL frameworks without relying on the availability of a trusted third party. This means the initialisation task is either performed by the FL server or FL client(s).

In~\cite{madi2021secure}, the initial keys vital for homomorphic encryption are generated by one of the clients. This client is either randomly selected by the server, or elected by clients following a given voting protocol. However, this approach only works under the assumption that clients are honest and the server does not collude with any client. Otherwise, a malicious server can select a colluding client to compromise with the key generation process, which is essential for the subsequent training and verification steps. For the verification algorithm, the authors leverage the Paillier cryptosystem~\cite{struck2017linearly}, a linearly homomorphic authenticated encryption scheme, to check the correctness of aggregated model returned by the server. However, the computational overhead of Paillier encryption becomes very high when the number of participants involved is large. 

Alternatively, VERSA~\cite{hahn2021versa} uses the public key infrastructure (PKI)~\cite{maurer1996modelling} to achieve safe initialisation without any trusted entity. PKI binds public keys with the respective participants when they advertise their keys to join the FL protocol. It reduces the computational cost by leveraging a lightweight pseudorandom generator to verify the aggregated FL model from the server. 

\subsubsection{Verifying the FL Clients}


The FL clients can also misbehave and deviate from the FL protocol. There are a few types of client misbehaviours. Firstly, clients can lie about the quantity and quality of their local data, which can mislead the FL model aggregation algorithms at the server. Besides, a client can forge its identity. These two misbehaviours are both about a client's local information. Secondly, clients who aim to free-ride on others and save their local computational resources can reuse local model updates from past iterations or even randomly generate model parameters to avoid actually performing local model training. Malicious clients can even poison the local model parameters in an attempt to compromise the final FL model. Therefore, it is crucial to verify clients' local information and their model updates. 
In this case, FL clients are provers and the FL server is the verifier. Works in this domain commonly make the following assumptions: 
\begin{enumerate}
    \item \textit{Semi-Honest FL Server}: The FL server is assumed to follow the FL protocols faithfully, but is curious about the clients' private information.
    \item \textit{Misbehaving FL Clients}: It is assumed that there exist FL clients who misbehave. The specific types of client misbehaviours vary among different works.
\end{enumerate}

\paragraph{Verifying Client Local Information:}
A client's local information consists of its identity and local data. The verification of client identity mainly relies on signatures or proofs generated by the clients. VERSA~\cite{hahn2021versa} leverages digital signatures to verify the identities of clients, making it applicable for scenarios in which the clients are malicious rather than just being honest-but-curious. Alternatively, the server authenticates client identity utilising a Fiat-Shamir zero-knowledge proof~\cite{shoup2000practical} before the start of the training process~\cite{mou2021verifiable}. 

Verification of a client's local data mainly relies on a safe data structure to store the data version information without exposing private information. In~\cite{zhang2020blockchain}, a blockchain is adopted to verify the integrity of client data. Each client periodically creates a Merkle tree in which each leaf node represents a client data record, and stores the tree root on the blockchain. When a dispute happens, the FL server can verify client data integrity by comparing it with the respective Merkle tree root stored on the blockchain. In~\cite{lo2021blockchain}, the hashed global model parameters and local data version information (including timestamps and data size) are recorded in a blockchain, enabling verification and auditing of client identity and local model performance. These two works focus on making useful information available for verification purposes, rather than how to verify the information.

The quality of a client's local data can be verified by influence function during the training process. The influence-based hierarchical framework~\cite{li2021efficient} enables the FL server to first identify the most negatively influential FL clients and find the most negatively influential data samples. By applying a participant selection strategy to remove these negative samples, the framework can fix bugs as well as accelerate FL model convergence. Similarly, in \cite{zhang2021incentive}, the data quality is verified using the loss on the validation set. If the difference between the loss of the global model formed with and without the suspicious local model exceeds a threshold, this model is discarded. Note that when verifying local data quality, clients are assumed to correctly follow the FL protocol and upload their models truthfully.

\paragraph{Verifying Client Model Updates:} 
During FL training, clients are supposed to train local models with their local data and upload the local model parameters to the server for aggregation. Verification during this process is essential for ensuring that clients follow this agreed protocol. 

Similar to verifying server aggregated models, the proof generated by the client is also used to verify the correctness of client uploaded model updates. TrustFL~\cite{zhang2020enabling} leverages trusted execution environments to verify if local FL model training has been executed with high confidence. Each client performs local model training in an isolated environment and outputs the model updates with its signature. The server verifies the signatures of randomly selected clients in each round of training to check their correctness. To prevent clients from contributing only when requested to be verified, TrustFL adopts a ``commit-and-prove'' design, requiring the clients to send a commitment message indicating the training rounds that they have finished before they know about whether they will be verified. The verification can later be conducted on a random claimed round. To mitigate misbehaviours involving reusing previously obtained local models to update the server, TrustFL devises a ``dynamic-yet-deterministic'' data selection method by requiring data owners to ensure that training data used for local model training are different at each round. As a result, model updates from the same data owner in different iterations shall be different.

Reputation schemes have been applied to help verify FL clients' contributions to the aggregated model. There are multiple ways to calculate a client's reputation, including multi-weight subjective logic~\cite{kang2020reliable}, historical contribution evaluation based on Shapley values~\cite{liu2021gtg,Liu-et-al:2022}, gradient vector similarity~\cite{xu2020reputation}, interaction behaviours~\cite{zou2021reputation}, test performance~\cite{wang2020novel,wang2021reputation}, beta distribution~\cite{song2021reputation} and subjective evaluation by FL task initiators~\cite{zhang2021incentive}. Besides evaluating client training contributions, reputation schemes have also been used to detect malicious users~\cite{song2021reputation}. Reputable clients have a higher probability of joining FL model aggregation and the next training iteration. Reputation values can be stored on a blockchain for efficient and transparent management~\cite{kang2020reliable,ur2020towards,zhang2021incentive,zhang2021blockchain}.

Contract theory has also been applied to verify FL clients' contributions~\cite{kang2019incentive,ye2020federated,lim2020dynamic,ding2020optimal,tian2021contract}. The FL server designs specific contracts for different types of data owners characterised by different levels of data quality and the corresponding payoffs. Each data owner must select and sign one of these contracts in order to participate in FL. In some works~\cite{kang2019incentive}, if the data owner cannot fulfil the expected contractual obligation, the FL server can adjust its reputation accordingly and withhold payment.


\subsection{Verifiable Decentralised FL Systems}
Decentralised FL systems do not require an FL server. The tasks of model aggregation and distribution are mainly performed by participating data owners. Blockchains are often leveraged by such FL systems due to their desirable qualities, including:
\begin{enumerate}
    \item \textit{Security}: Once a block is accepted and added to the blockchain, no one can tamper with the records without being detected. This property makes blockchain secure storage for ensuring information integrity.
    \item \textit{Automation}: The smart contracts in blockchains ensure that once a pre-defined condition is met, specific operations can be triggered. This automatic contract checking scheme enhances the efficiency of blockchains.
    \item \textit{Transparency}: Information stored on the blockchain (e.g., FL model updates) is accessible by all, enabling them to perform verification and auditing conveniently.
\end{enumerate}


In blockchain-based decentralised FL systems, the trainers (i.e., data owners) refer to the smart contract to train their local models and upload their local parameter updates to the blockchain network. Selected participants or organisations can register as workers (a.k.a. miners) in the blockchain, who take charge of verifying the information uploaded by trainers. Each worker mines according to the blockchain consensus protocol until it can generate a new block or a block is received from another worker. When a worker generates a new block, the other workers verify the contents of the block (e.g., random numbers, the status of smart contract changes, aggregated models). If a block is agreed upon by the majority of the workers, the block is added to the blockchain and accepted by the network. After the new block containing the updated FL model is added to the blockchain, all participants can update their models according to the information stored and begin the next round of FL model training. 

\subsubsection{Verifying the FL Trainers}
In decentralised FL settings, trainers upload model updates to the blockchain. Verification of FL model updates occurs before workers can aggregate them. In this scenario, trainers are the provers and each worker is a verifier. The common assumption made by existing works is that:
\begin{enumerate}
    \item \textit{Dishonest Trainers}: A trainer might upload forged model parameters to the blockchain.
\end{enumerate}

In BlockFL~\cite{kim2019blockchained}, a blockchain is designed for exchanging local model updates, verifying the submitted information, and allocating rewards. Specifically, workers in BlockFL are either randomly selected devices or separate nodes such as network edges. Each device engaging in the FL protocol is assigned to a random worker. Workers verify the received model updates before aggregation. They compare the sample size of a device with the time taken for it to produce the model updates to identify misbehaving devices which attempt to avoid actually training the local model. 

Proofs are also used to verify trainers' behaviours. In~\cite{bao2019flchain,zhao2020mobile}, model gradients generated by the trainers are masked and uploaded with their correctness proofs. The workers, who are registered trainers, check the masked gradients based on the proofs to determine which of them shall be selected for model aggregation. The selected gradients are then aggregated by the workers following the blockchain consensus protocol. A worker is selected as the leader based on verifiable random functions~\cite{zhao2020mobile} or its reliability ranking~\cite{bao2019flchain} to document its aggregation results into the blockchain. The reliability reflects each trainer's performance. It jointly considers the proof verification results, the model buyer's evaluation and whether it has successfully identified misbehaviours (if the trainer has been selected as a worker). A trainer's reliability value will also impact its chance to participate in future FL model training. Besides, the systems support third-party auditors to investigate misbehaviours based on the training records in the blockchain. 

\subsubsection{Verifying the FL Workers}
Workers in a blockchain are responsible for FL model aggregation. The verification of FL model aggregation takes place after a worker has successfully aggregated the received local model gradients. The system needs to verify whether the worker has performed the aggregation correctly. In this scenario, the worker who adds the new block to the blockchain is the prover and other workers are the verifiers. The following common assumption is made by existing works:
\begin{enumerate}
    \item \textit{Unreliable Workers}: It is assumed that there exist compromised workers who might not perform FL model aggregation following the correct protocol, while the majority of the workers remain trustworthy.
\end{enumerate}

In VFChain~\cite{peng2021vfchain}, each trainer generates a signature and upload it to the blockchain together with its model updates. The workers compete with each other to create a valid block containing the aggregated FL model and signatures from the trainers. A trusted committee is formed by randomly selected workers to verify the aggregated model according to verifying contract. The committee is continuously updated over training iterations. Historical aggregated models, together with verification signatures, are stored in the blockchain which is immutable to modification and accessible to all users. To improve the auditing efficiency, an authenticated data structure has been proposed based on a dual skip chain to store the training trails. Nevertheless, VFChain can only work when the trainers are semi-honest and do not collude with each other. 


\section{Verifiable FL Evaluation Methods}
To evaluate the effectiveness of a verifiable FL approach, a combination of theoretical analysis and experimental evaluation is commonly adopted.

\subsection{Theoretical Analysis of Verifiability}
Theoretical analysis of verifiability is most relevant for FL verification approaches designed to detect misbehaviours by FL participants~\cite{xu2019verifynet,kang2019incentive,bao2019flchain,guo2020v,zhang2020privacy,fu2020vfl,zhang2020enabling,jiang2021pflm,madi2021secure,hahn2021versa,song2021reputation,mou2021verifiable,han2021verifiable}. These works generally require the prover to generate a proof, a signature or a ciphertext for the verifier to check. The goal of theoretical analysis is to show that, under the given verifiable FL scheme, once the generated proof is verified to be correct by the verifier, it can be concluded that the activity being verified is correct. 

The specific approach to proving the verifiability of a given scheme depends on the proof generation technique used. Taking the works based on homomorphic hash functions and pseudorandom functions to verify the correctness of aggregation from the server for example~\cite{xu2019verifynet,mou2021verifiable,han2021verifiable}, these works perform analysis based on the l-BDHI assumption~\cite{fiore2014efficiently} and the DDH assumption~\cite{boneh2001identity} to prove that the proposed scheme can conclusively verify the prover's activities. Verifiability analysis is a useful approach to show the theoretical foundation and expected performance bounds of given verifiable FL approaches.

\subsection{Experimental Evaluation Metrics}
Experimental evaluations are useful to test given verifiable FL approaches in environments with simulated or actual misbehaviours in order to study their effectiveness in complex settings. The following evaluation metrics are commonly used to quantify the effectiveness of these approaches.

\paragraph{Model Performance:}
Model performance is an important evaluation metric for verifiable FL systems as the goal of FL participants is to obtain an improved model. In FL settings, model performance is closely related to the number of participating data owners and the size of the local gradient updates from them. Experiments often fix the number of participating data owners while varying the size of their local gradient updates (and vice versa) to study the FL model performance. Test accuracy is commonly used to evaluate model performance. Precision, recall and F1-score have also been used to evaluate verifiable FL approaches designed for classification models~\cite{zhang2020blockchain}. 

\paragraph{Verification Overhead:}
Verification incurs computation and communication overhead, which is related to the number of data owners and the size of the local gradient updates. Experiments designed to measure computation and communication overhead are often conducted in settings where these two factors are varied over a range of values. 
The computation cost is quantified by the time taken for verification. It can be separated into the time taken for proof generation and validation~\cite{han2021verifiable}, or in finer granularities (e.g., time taken for key generation, encryption, verification, decryption and evaluation~\cite{madi2021secure}). One-time computation costs (e.g., time for secure setup) are often ignored. 
Communication overhead is measured by the total amount of data transmitted during the verification process. This includes the amount of transmission required to successfully upload the verification information onto the blockchain in blockchain-based verifiable FL~\cite{jiang2021pflm,lo2021blockchain}.



\paragraph{Resistance to Attacks:}
To evaluate the robustness of verifiable FL, resistance to attacks has been used as an indicator. These attacks are conducted in simulated environments. In works designed to verify the training performance of FL clients \cite{kang2020reliable,kang2019incentive}, the quality of local training data, the number of attackers and the intensity of attacks are varied to simulate different conditions in which the verifiable FL approach needs to operate. The performance of the FL model produced in such conditions is used to measure how resistant the approach is in the face of attacks.

\section{Promising Future Research Directions}
Through our survey, it can be observed that research in verifiable FL today mostly focuses on making key information available to facilitate verification, and performing verification on specific activities and roles in various types of FL systems. To advocate verification to become an integral part of future FL systems to allow the community of participants to self-police and self-organise, we envision the following future research directions.

\paragraph{Relaxation of Simplifying Assumption:}
Current works commonly assume at least one party in FL is semi-honest. In centralised FL, if clients want to verify the correctness of the aggregated model returned by the FL server, all clients are assumed to be semi-honest. Similarly, the FL server also needs to be semi-honest when it verifies the clients. This makes existing approaches vulnerable in situations in which the server and clients are malicious or collude. This limiting assumption needs to be relaxed to enable future verifiable FL approaches to handle more challenging real-world conditions.

\paragraph{Verification without Trial FL Model Training:}
In FL settings, participants often need to self-declare certain information in order to be selected to join a federation. Such self-declared information includes the quality and quantity of local data, and the computational resource of a data owner. Due to privacy concerns, the verifier cannot directly inspect such information to check if it is truthfully declared. Thus, existing verifiable FL approaches generally require candidate data owners to participate in FL model training first, before they can verify their claims on local resource commitment by observing their performance during training. To reduce the waste on computation and communication resources as a result of this trial-and-error approach, research on privacy-preserving verification of self-declared information without requiring trial participation in FL model training is desired.

\paragraph{Automating Auditing Report Generation:}
Decentralised verifiable FL approaches often keep traceable trails of verification for auditing purposes. These records usually contain model updates, aggregated models and verification proofs, etc. However, raw records are not convenient for third-party auditors to make sense of. This source of information can be leveraged in combination with explainable AI (XAI) research \cite{Tjoa-Guan:2021} to automate the generation of auditing reports to enhance the transparency of the FL verification processes. By dynamically incorporating XAI techniques \cite{Kim-et-al:2021} such as explicit explanation, alternative explanation, knowledge of the explainees and interaction designs with FL verification, auditing reports for diverse FL stakeholders with different needs for insight can be generated, thereby enabling wider participation in upholding the integrity of verifiable FL.

\paragraph{Verification-based Community Sanctioning:}
Identifying individual instances of misbehaviour by FL participants is important. Nevertheless, in order to build long-term insight into participants' behaviour patterns, it is necessary to investigate how to leverage the verification results to form the basis of community sanctioning. Preliminary works such as \cite{li2021efficient} are starting to emerge in this area which attempts to build trust among participants based on verification results. Research in verification-based social norm formation \cite{Shulman-et-al:2017} and interaction strategies can enable FL systems to self-police their members and deter misbehaviours.

\bibliographystyle{named}
\bibliography{ijcai22}
\balance

\end{document}